# High-temperature thermal conductivity measurements of macro-porous graphite

Shomik Verma[1], Michael Adams[2], Mary Foxen[1], Bryan Sperry[1], Shannon Yee[2], Asegun Henry[1,*]

[1] Massachusetts Institute of Technology, 77 Massachusetts Avenue, Cambridge, MA 02139
[2] Georgia Institute of Technology, 495 Tech Way NW, Atlanta, GA 30318

## ABSTRACT

Graphite is a unique material for high temperature applications and will likely become increasingly important as we attempt to electrify industrial applications. However, high-quality graphite can be expensive, limiting the cost-competitiveness of high-quality graphite technologies. Here, we investigate the thermal properties of low-cost, low-quality, macro-porous graphite to determine the tradeoff between cost and thermal performance. We use laser flash analysis (LFA) to measure the thermal diffusivity of graphite at high temperatures. However, due to the large pores in the graphite samples preventing uniform laser flash heating, we must apply a thick coating to achieve the required flat, parallel surfaces for LFA measurements. The presence of the coating directly impacts the measured diffusivity, not only because of the added thickness but also because of the sample/coating interface profile generated. We therefore develop a methodology based on finite element modeling of a variety of sample/coating interface profiles to extract properties of the sample. Validating the methodology against a reference sample demonstrates a mean absolute percentage error of 8.5%, with potential improvement with better sample characterization. We show low-cost, low-quality graphite has a thermal conductivity of ~10 W/m/K up to 1000 °C, which is an order of magnitude lower than high-quality graphite, but contributions from photon conductivity may result in higher conductivities at higher temperatures. Overall, we demonstrate an approach for measuring thermal properties of macro-porous materials at high temperatures, and apply the approach to measuring thermal conductivity of porous graphite, which will aid in the design of high-temperature systems for cost-competitive decarbonization.

## 1. INTRODUCTION

Graphite is used in many high-temperature applications including as heating elements for furnaces,[1,2] storage media and containment vessels for thermal energy storage,[3,4] or moderators for nuclear reactors.[5,6] Many of its properties make it ideal for these high-temperature applications, including no melting at atmospheric pressures, sublimation at >3000 °C, stability in non-oxidizing atmospheres, and high electrical conductivity.[7,8] These applications of graphite will become increasingly common as the world seeks to decarbonize industrial heating and deploy more renewable energy generation. Using lower-cost graphite can reduce capital costs and help make electrified technologies cost-competitive.[9]

However, low-cost graphite is often low-quality, with high porosity and large pores. This makes understanding its thermal properties difficult. First, the properties of low-quality graphite differ from bulk graphite due to the presence of these pores, so the properties of bulk graphite cannot be assumed. Further, the large pores make measurement of thermal properties difficult, as many techniques require flat surfaces.

*Corresponding Author: ase@mit.edu

This study aims to measure high-temperature thermal properties of low-quality graphite using the laser flash analysis (LFA) approach for experimental measurements. As the LFA requires flat, parallel surfaces for accurate measurement,[10] the graphite samples are first coated with a putty material. Effective medium simulations are then used to extract the desired properties of low-quality graphite from the measured values.

Previous studies have measured thermal conductivity of high-quality graphite.[11–13] For example, Acheson graphite has been extensively studied for many decades.[14] Of specific interest to our work, McEligot et al. measured thermal conductivity of G-348 fine-grained isostatic graphite using LFA, differential scanning calorimetry (DSC), and dilatometer (DIL) measurements.[15] Das et al. measured and modeled conductivity of microscale exfoliated graphite and explicitly included the contributions of photon conductivity at high temperatures.[16]

When measuring thermal properties, LFA samples are often coated with a thin graphite layer to better absorb the heating pulse.[17] Many previous studies have investigated the impact of this coating on the measured properties.[18–21] Rapp investigated the impact of $5\mu$m graphite coatings on glass-epoxy composite samples for electronic applications.[22] From a theoretical standpoint, Blumm et al. developed an analytical model for the transient temperature profile of 3-layer materials (coating – sample – coating) given an heat pulse at one end.[23] Lim et al. determined the error in LFA measurements if the coating layers are neglected, concluding errors up to 33% are possible.[24]

This work fills two essential gaps in the literature. First, we measure properties of macro-porous, low-quality graphite that has not been investigated previously. Second, most previous works studying the impact of coating LFA samples assume a flat interface between sample and coating. However, this is likely not the case for samples with large pores. One previous work investigated surface topology effects on apparent diffusivity, but was limited to a few simple surface geometries that may not be physically similar to samples.[25] Therefore, our work is unique in explicitly considering the rough interface profile between sample and coating and determining how to accurately back-calculate the desired thermal property of graphite from the measured property.

## 2. METHODS

### 2.1. Experimental

*Sample preparation.* Large samples were obtained from various vendors at different levels of quality. For this study, two grades of graphite (China Rongxing Company) were used. First was low-quality graphite formed from recycled graphite products with other raw materials, costing approximately $0.70/kg. We also obtained medium-quality graphite formed from machining scrap and powder, which was approximately $1.50/kg. Fig 1(a) shows the low quality (S2), medium quality (SM), and higher grades (30%, 50%) of graphite for comparison.

LFA requires samples with 10-12 mm diameter and 3-5 mm thickness, so the bulk materials were machined to 10 mm OD by 3 mm thickness samples. Following this step, the mid-quality graphite had flat, parallel surfaces, as seen in Figure 1(b) and (g), and its thermal properties could be measured directly. However, most samples of the low-quality graphite were significantly rougher, as seen in Figure 1(c), except for one sample, shown in Figure 1(d). Because only one low-quality sample was able to be measured without coating, we do not have a measurement of uncertainty, and additionally we do not have a sense of variability of diffusivity in the sample itself. To attain higher certainty in our results for low-quality graphite, we coat the other samples, as shown in Figure 1(e), allowing us to measure their thermal properties.

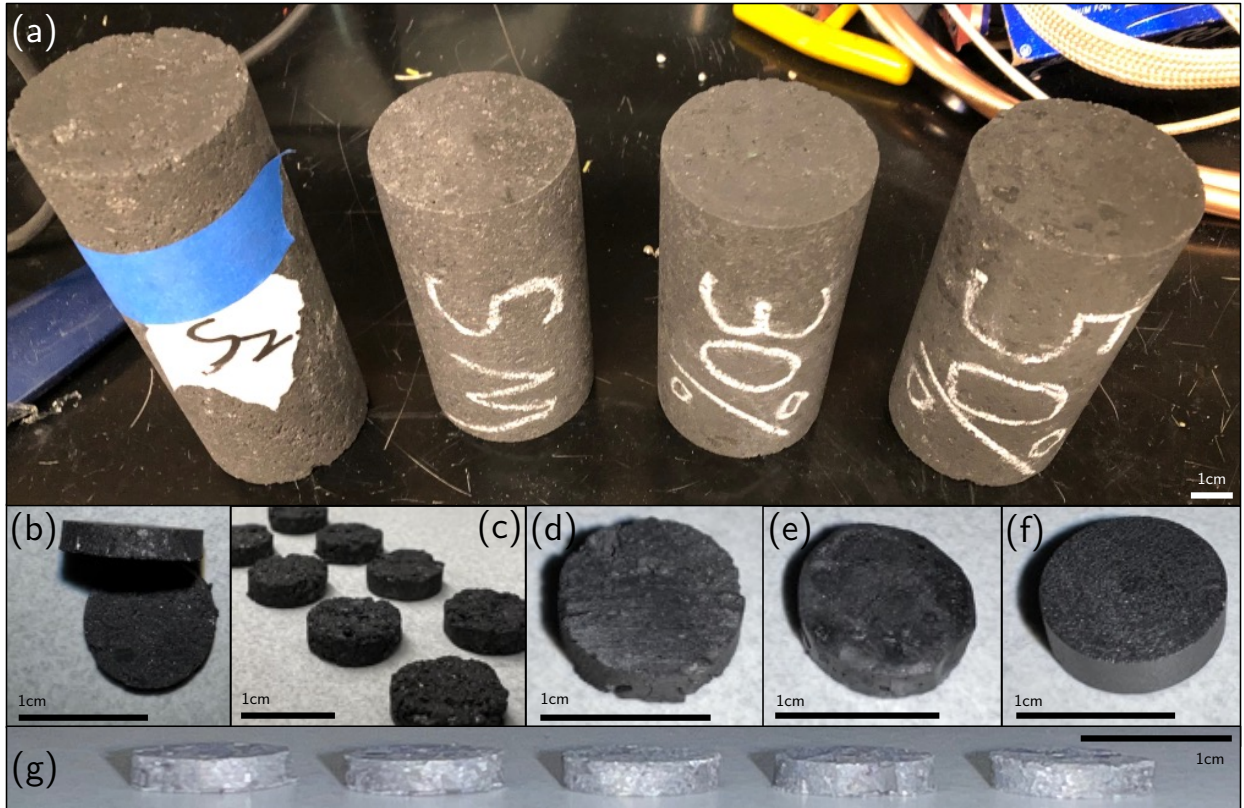

**Fig 1**: (a) Bulk samples of various grades of graphite received from China Rongxing Group. S2 formed from regeneration of recycled graphite and raw material, SM formed from machining scraps, and 30%/50% are conventionally molded graphite blocks. (b,g) Samples of medium-quality graphite (SM) machined to LFA dimensions. (c) Uncoated low-quality graphite (S2) sample machined to LFA dimensions. (d) Reference sample for uncoated low-quality graphite which did not need coating for measurement. (e) Coated low-quality graphite sample with putty coating. (f) EBS-1X putty coating sample prepared for LFA measurement.

The coating process was as follows. EBS-1X graphite epoxy (CeraMaterials) was used as the putty coating material. First, a uniform layer of thickness approximately 1mm was applied on one side of the material to ensure all pores were sufficiently filled. Then, the sample was baked at 130 °C for 2 hours. Next, the hardened coating was machined, and the thickness and mass added was measured. This procedure was repeated for the other side. The resulting part is shown in Figure 1(e), and clearly has a much smoother surface compared to the samples in Figure 1(c). Samples of the putty were also made for inputs to effective medium theory calculations and are shown in Figure 1(f). The dimensions and naming of the samples considered in this study are shown in Table 1.

**Table 1**: Sample dimensions and masses for low-quality graphite samples. Note that one low-quality (sample 5) sample was uncoated. Five mid-quality samples (not included below) were all $3 \pm 0.03$ mm thickness x $10 \pm 0.03$ mm diameter, and weighed $0.376 \pm 0.003$ g. Sample 3 was only coated on the top and bottom while samples 2 and 8 were also coated on their sides.

| Sample number | Initial thickness (mm) | Initial diameter (mm) | Final thickness (mm) | Final diameter (mm) | Initial mass (g) | Final mass (g) | Added mass (g) |
|---|---|---|---|---|---|---|---|
| 5 | 3.11 | 10.15 | - | - | 0.359 | - | - |
| 3 | 3.11 | 10.10 | 3.19 | 10.10 | 0.349 | 0.394 | 0.045 |
| 2 | 3.12 | 10.12 | 3.19 | 10.80 | 0.350 | 0.447 | 0.097 |
| 8 | 3.10 | 10.05 | 3.40 | 11.20 | 0.371 | 0.505 | 0.134 |

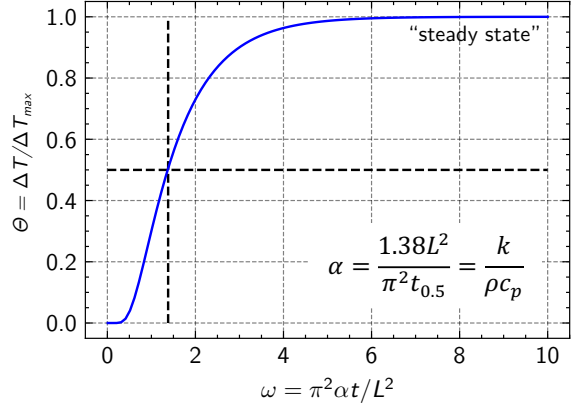

**Fig 2**: Representative temperature profile expected from LFA operation. Applying a fast laser pulse on one side of the material produces this temperature response on the other side. Non-dimensionalizing time and temperature allows calculation of thermal diffusivity.

*Measurements.* The LFA was chosen for measuring thermal properties in this study due to its high-temperature capabilities. While some LFAs have capabilities up to 2800 °C,[26] the LFA furnace used in this study (NETZSCH LFA 427) had a limit of 1200 °C. The LFA requires a relatively thin (1-5 mm) cylindrical sample with diameter 10-12 mm. The surfaces of the sample must be flat and parallel, to ensure good thermal contact with the sample holder. The sample should also be opaque with high absorptivity and emissivity to ensure the heating and sensing function properly.[10]

The LFA operates by heating the bottom of the sample with a fast laser pulse. The temperature response at the other side of the sample is detected with an IR sensor, and this transient response is translated to a diffusivity. Figure 2 shows the non-dimensional temperature response and time.[27]

From this non-dimensionalization, we can extract the thermal diffusivity of the sample ($\alpha$) by measuring the time it takes for a sample of thickness $L$ to reach half of its maximum temperature ($t_{0.5}$), using the following equation:

$$\alpha = \frac{1.38 L^2}{\pi^2 t_{0.5}}$$
(1)

This apparent $\alpha$ (mm²/s) can then be used to calculate thermal conductivity with the relation:

$$\alpha = \frac{k}{\rho c_p} \tag{2}$$

where $k$ (W/m/K) is the thermal conductivity, $\rho$ (kg/m³) is the density, and $c_p$ is the specific heat (J/g/K). It naturally follows that the specific heat and density of the sample must also be measured. We use differential scanning calorimetry (DSC, NETZSCH) to measure the specific heat of the samples. We average DSC measurements over 3 samples for uncertainty quantification. For density, we measure mass and volume of the bulk material at room temperature (1.6 g/cm³ for graphite and 1.4 g/cm³ for putty) and extrapolate to higher temperatures using a linear expansion coefficient.[15] For more accurate results, a TMA or DIL could be used to directly measure thermal expansion as a function of temperature.

LFA and DSC measurements were completed on three medium-quality (uncoated) graphite samples, one uncoated low-quality sample, three coated low-quality samples, and three putty samples.

For the uncoated and putty samples, calculating the conductivity is straightforward as only one material is present in the sample. For the coated samples, however, an effective medium theory must be used to back-calculate the desired properties from the measurements. Therefore, measurements of the putty coating must be conducted. Namely, the thickness and mass of the sample with and without coating was measured, which allows calculation of the original graphite volume and added putty volume. We also measure the thermal properties of the putty, including $c_p$, $\rho$, and $\alpha$, which means we know $k$ of the putty, and we also measure $c_p$ and $\rho$ of the graphite sample. However, there are many unknowns that preclude direct extraction of sample $\alpha$ and $k$ from these knowns.

Particularly, the profile of the graphite/putty interface is unknown. This is a function of the porosity of the graphite and the distribution of pore sizes along the surface. Understanding this requires microstructure characterization of the graphite using 3D imaging methods such as micro computed tomography. However, we are interested in determining how important these interface profiles are, and whether we can develop an effective medium theory that can determine the graphite properties with reasonable accuracy without requiring expensive imaging. We thus turn to simulations.

## 2.2. Computational

As suggested above, we are interested in evaluating the sensitivity of effective thermal properties on the graphite/putty interface profile. Therefore, we conduct finite-element simulations of a variety of interface profiles to determine the variability in effective properties. The 1D nature of the LFA operation (heat diffuses from one end of the sample to the other) suggests that we can simplify our problem from 3D to 2D and use area fractions instead of volume fractions. Through our measurements, we know the added area of the putty. Thus, our goal is to generate many interface profiles with the same area but different geometries. We employ the following workflow for this.

First, we generate normalized interface profiles, which are scalable to different geometries. We decide on a nominal geometry of 3 mm thickness and 10 mm diameter. Due to the presence of pores on the surface, the initial area of the graphite will be less than the nominal 30 mm². We then decide on a "missing" area of 10% of the nominal area, meaning 3 mm² of the area will be occupied by putty instead of graphite. Then, we generate various interface profiles that fit this area.

We employ a Monte Carlo method to generate these profiles. First, the 10 mm diameter is discretized in the *x*-direction into a certain number of points, e.g. 10. For each of these points, a *y*-coordinate is chosen by randomly sampling between 0 and the maximum depth of the pore. We are specifically interested in the difference in having many small pores or few large pores, so this maximum depth is varied from 10% of the overall thickness to 20%. A spline is then created between these points. If the area under the spline matches the desired area, the curve is chosen. The procedure is repeated for both sides of the part. Figure 3(a) shows a schematic of this interface profile generation technique, and Figure 3(b) shows an example of a profile with a maximum depth of 10% of the thickness, while Figure 3(c) shows a profile with a maximum depth of 20% of the thickness. A total of 100 interface profiles are generated with this methodology.

Once the interface profiles are generated, they can be imported into a finite element software for simulation. In this case we use COMSOL for heat transfer simulations. For transient conduction, we simulate the LFA by applying a heat flux of 100 W/cm² for 1ms to the bottom of the sample. We record the transient temperature at the top surface of the sample, and the apparent diffusivity can then be calculated from Equation 1 above.

Thus, through these simulations we can determine if knowing the exact interface profile is necessary to extract the desired sample properties.

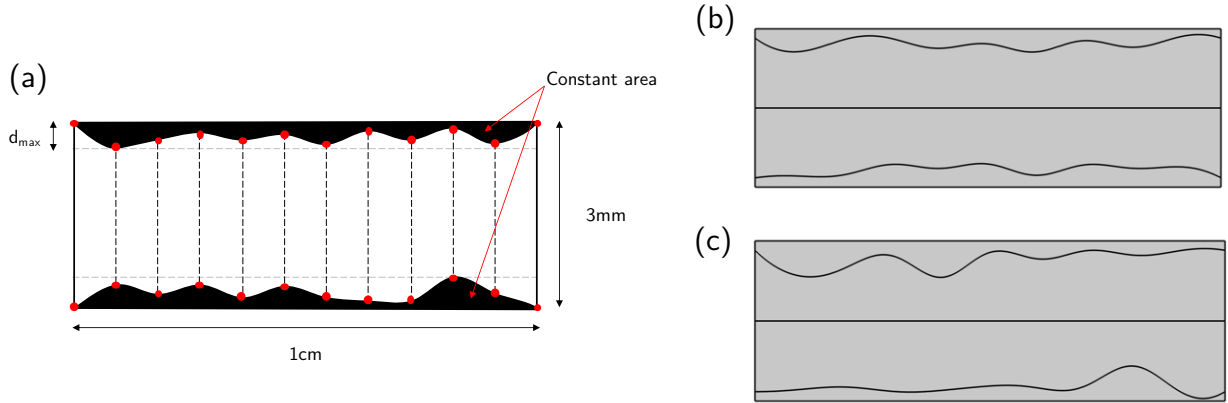

**Fig 3:** (a) Schematic of surface profile generation methodology. 10 points evenly spaced in x are chosen to start. Then, the y-coordinates are selected randomly between the set bounds. A spline is then created, and if the area of the spline matches the desired area, it is selected. The bounds can be varied as desired, as shown in (b) which has a maximum depth ($d_{max}$) of 0.03mm while (d) has a maximum depth of 0.06mm.

## 3. RESULTS AND DISCUSSION

### 3.1. Experimental Measurements

First, we measure properties of the mid-quality graphite which had flat, parallel surfaces suitable for LFA measurements. These are shown in Figure 4 and Figure S9(a). The specific heat, shown in Figure S9(a), was measured to 500 °C with three samples and a polynomial fit following the form of Butland and Madison[28] was used to extrapolate to higher temperatures:

$$c_p = a_1 + b_1 T + c_1 T^{-1} + d_1 T^{-2} + e_1 T^{-3} + f_1 T^{-4} + g_1 T^{-5} \qquad (3)$$

Thermal diffusivity was measured to 1200 °C with three samples, as shown in Figure 4(a), and a similar polynomial fit was used with an additional term to account for photon diffusion[16] at higher temperatures:

$$\alpha = a_2 + b_2 T + c_2 T^{-1} + d_2 T^{-2} + e_2 T^{-3} + f_2 T^{-4} + g_2 T^{-5} + hT^3 \qquad (4)$$

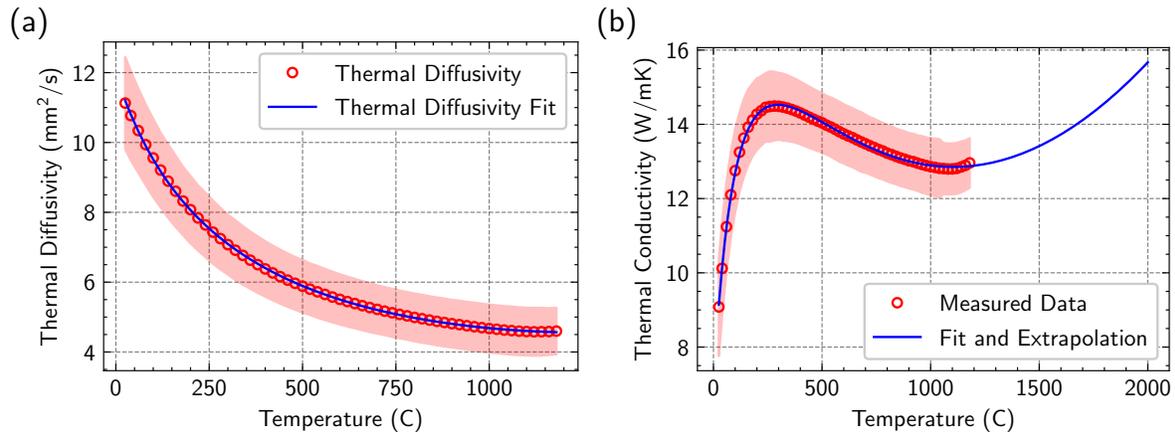

**Fig 4:** (a) Specific heat, (b) thermal diffusivity, and (c) calculated thermal conductivity for medium-quality graphite samples, with 95% confidence bounds. Fits are polynomial in nature as described in Equations 3 and 4.

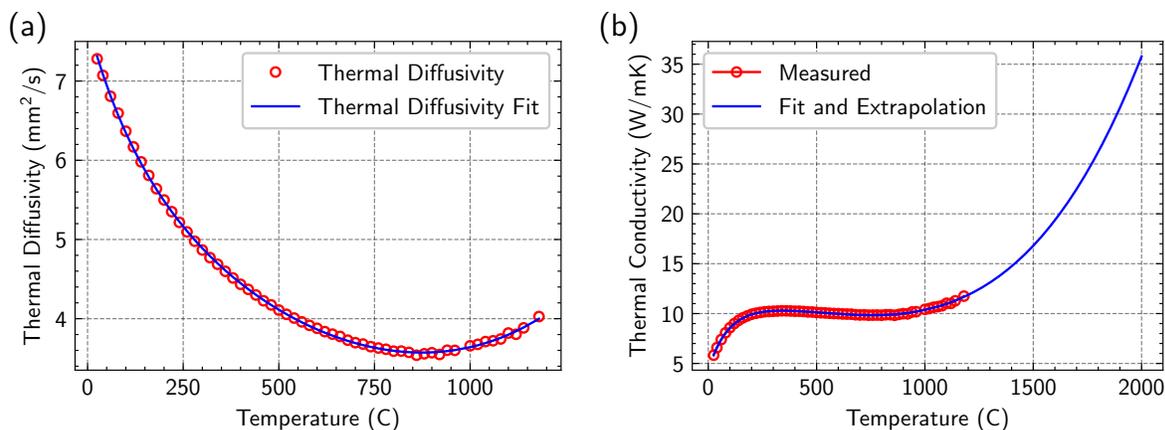

**Fig. 5**: (a) Specific heat, (b) thermal diffusivity, and (c) calculated thermal conductivity for low-quality graphite samples, with 95% confidence bounds where possible. Fits are polynomial in nature as described in Equations 3 and 4. 3 samples were used for specific heat but only 1 sample was available for diffusivity measurements.

Finally, the thermal conductivity can be calculated from Equation 2 above and is shown in Figure 4(b). Using the two polynomial fits above, we can extrapolate conductivity to even higher temperatures, and note that the conductivity increases due to the contribution of photon conductivity. However, the exact trend may be inaccurate as limited data is available above 1000 °C.

Next, we measured properties of the one low-quality sample with flat surfaces. We conduct similar fits for specific heat (Figure S9(b)) and diffusivity (Figure 5(a)) for this new sample, and then fit and extrapolate thermal conductivity. From this, we see that this low-quality graphite has lower thermal conductivity at the lower temperature range, of ~10 W/m/K compared to ~13 for medium-quality graphite. This is expected as the lattice conductivity of low-quality graphite is lower due to its high porosity. However, at high temperatures, there may be a benefit to low-quality graphite due to the contributions of photon conductivity. This is apparent when comparing Figure 5(a) with Figure 4(a). For low-quality graphite the diffusivity starts to increase at ~800 °C while the medium quality diffusivity is still stagnating at 1000 °C. However, further measurements at high temperatures are required to confirm this.

We now turn to the coated low-quality samples. First, we measure the properties of the putty coating, which are shown in Figure S8. We also know the specific heat of the graphite sample from Figure S9(b). We thus lastly need the apparent diffusivity of the coated samples and should then be able to extract the diffusivity/conductivity of the sample from effective medium theory. Figure 6 shows the apparent diffusivity for three coated samples.

These measurements show similar trends as the uncoated graphite, with a decreasing thermal diffusivity that stagnates and then increases around 800 °C. The absolute value of the diffusivity is lower, which is expected as the diffusivity of the putty is lower, as shown in Figure S8(b).

Now that we have obtained all the data, we turn to effective medium theories derived from simulation to determine the desired thermal properties of the graphite sample.

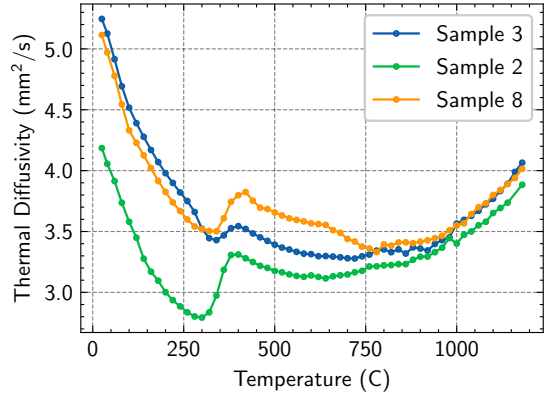

**Fig. 6:** Apparent thermal diffusivity for 3 coated samples. Sample 3 was only coated on the top/bottom, while samples 2 and 8 were additionally coated on the sides. The jump in properties at ~300 °C was likely due to the putty coating rather than the graphite sample itself, as seen when comparing the diffusivities of putty (Figure S8) and uncoated graphite (Fig. **5**).

### 3.2. Simulations

The LFA measures apparent thermal diffusivity of a sample, and because diffusivity is inherently a transient phenomenon, a resistance analysis cannot be used. The full heat equation must be solved, and while this has been done for simple geometries,[20] in our case, the complex interface profile makes an analytical solution impossible, and finite element methods (FEM) are required.

Thus, instead of formulating a closed-form effective medium theory, we use FEM to determine the thermal conductivity/diffusivity of the sample. Namely, we use the given properties (geometry, specific heat, density, putty conductivity) at each temperature value, input a certain graphite conductivity, and simulate the LFA operation in COMSOL for each of the 100 interface profiles generated. This generates 100 different apparent diffusivities. If the range of diffusivities includes the experimental diffusivity, the selected conductivity is a possible conductivity of the material. This analysis therefore outputs a range of possible conductivities for each measured diffusivity, providing a statistical distribution of possible values. The results of this analysis are shown in Figure 7.

Figure 7(a) shows the results for the three coated low-quality graphite samples considered in this work. We note that all 3 samples exhibit similar trends and have similar conductivity values as well. We can then average these values, with the purpose of getting a better understanding of the conductivity of the larger bulk material instead of small samples that may misrepresent the overall structure. The averaged thermal conductivity is shown in Figure 7(b) and is compared against the uncoated low-quality sample. The values match well, with a maximum error of ~15%, and the uncoated sample is always within the error bounds. We further note that Sample 3, which was only coated on the top and bottom, has the closest match to the uncoated Sample 5, indicating side coatings may negatively impact accuracy. This suggests the added putty volume should be kept to a minimum to achieve the best results. Regardless, the close match between uncoated and back-calculated conductivities serves as an important validation of our approach.

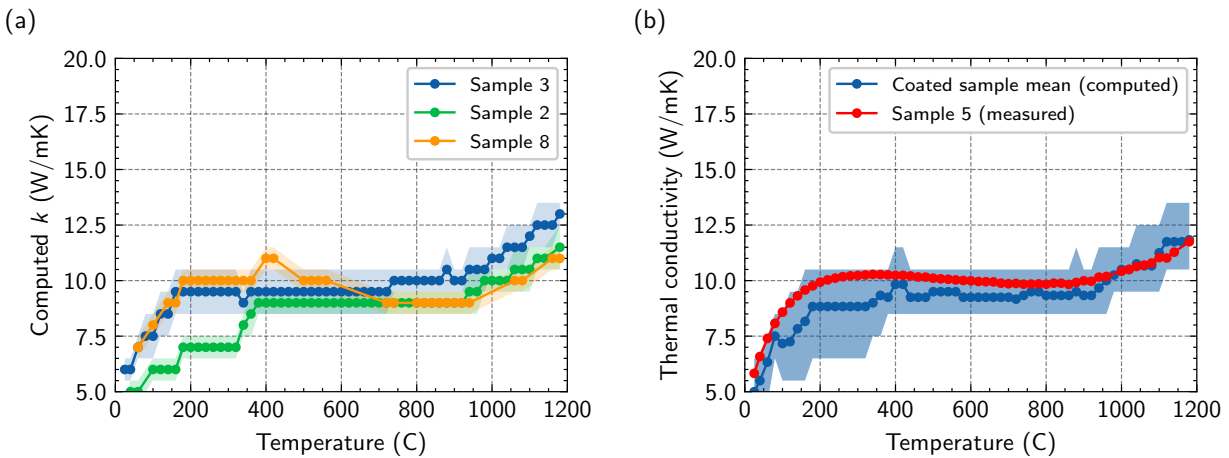

**Fig. 7:** Determining the conductivity of the graphite sample using FEM simulations. (a) Shows the mean and 95% confidence intervals for sample thermal conductivities for the 3 samples measured in Figure 6. (b) Averages the values of the 3 samples and compares against the reference sample.

Compared to conventional, high-quality graphite, we see that both mid- and low-quality graphite has significantly lower (~10x) thermal diffusivity and conductivity at room temperature up to 1000 °C.[15] However, extrapolations show that contributions from photon conductivity may be more significant in these macro-porous graphite samples, which could result in better performance than high-quality graphite at high temperatures. However, actual measurements at >1500 °C are required to confirm this.

## 4. CONCLUSIONS AND FUTURE WORK

Overall, in this work we have investigated the thermal properties of low-quality graphite. Because it is formed from recycled graphite or machining scraps, it often has high porosity and large pores, which impact its thermal performance. In this work, we directly measure thermal diffusivity of mid-quality graphite and then calculate thermal conductivity by knowing specific heat and density. For low-quality graphite, however, direct measurement of diffusivity is difficult, and only one sample had the flat, parallel surface properties necessary to complete the measurement. Thus, we introduce a coating procedure that goes beyond conventional thin LFA coatings to mask the macro-scale pores in the samples. In doing so, we must explicitly consider the sample/coating interface profile.

Therefore, we develop an effective medium theory methodology that uses finite element methods to model a variety of interface profiles and determine a range of possible conductivities for each measured diffusivity. These values are compared against the one uncoated sample, with good agreement (~8.5% mean absolute percentage difference), validating our approach. In addition to the accuracy, another benefit includes only requiring simple geometric and mass measurements and not 3D imaging of the interface.

There are several avenues of future work. Primarily, we are interested in further investigating the phenomenon of photon conductivity and therefore would like to take higher-temperature measurements of similar samples. Second, there were a few aspects not considered in the work: (i) We assumed the sample was impermeable to the putty coating, which was verified by thickness/volume measurements, but cross-sectional imaging could validate this assumption. (ii) Microstructure characterization with micro computed tomography could also help us better understand the primary heat transfer mechanisms and the impact of pores. (iii) This combined with computational tools such as PuMA could offer an alternative verification of measured results. (iv) We also did not directly measure density of the samples, so measuring thermal expansion as a function of temperature directly could help improve the accuracy of results.

Broadly, we hope this work offers a better understanding of the thermal properties of low-quality graphite. These properties could be used in design and simulation of high-temperature technologies to optimize their performance.

## ACKNOWLEDGEMENTS AND DATA AVAILABILITY

We would like to thank Alessandro Turchi and Bernd Helber for useful discussions on thermal conductivity and diffusivity measurement techniques. Data, code, geometry files, and COMSOL files are available on GitHub.[29]

# 6. SUPPLEMENTARY INFORMATION

## 6.1. Additional measurements

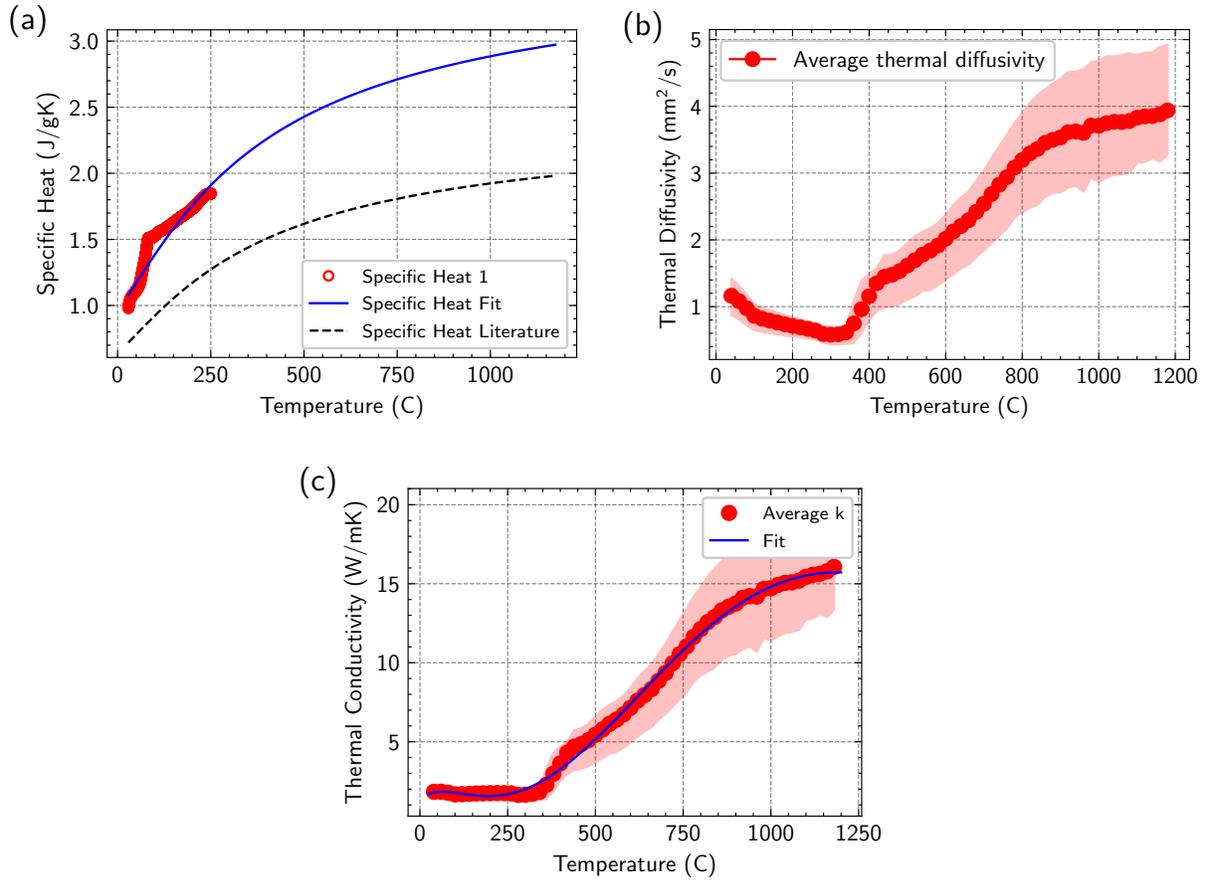

Figure S8: (a) Specific heat, (b) thermal diffusivity, and (c) calculated thermal conductivity for putty coating samples, with 95% confidence bounds where possible. Fits are polynomial in nature as described in Equations 3 and 4. 1 sample was used for specific heat and 3 samples were available for diffusivity measurements.

We note briefly here that the jump in properties in 300C may be due to the putty not being fully cured. Regardless, we expect the phenomena to be the same for bulk putty samples and putty coatings, so these results are applicable to our effective medium analysis.

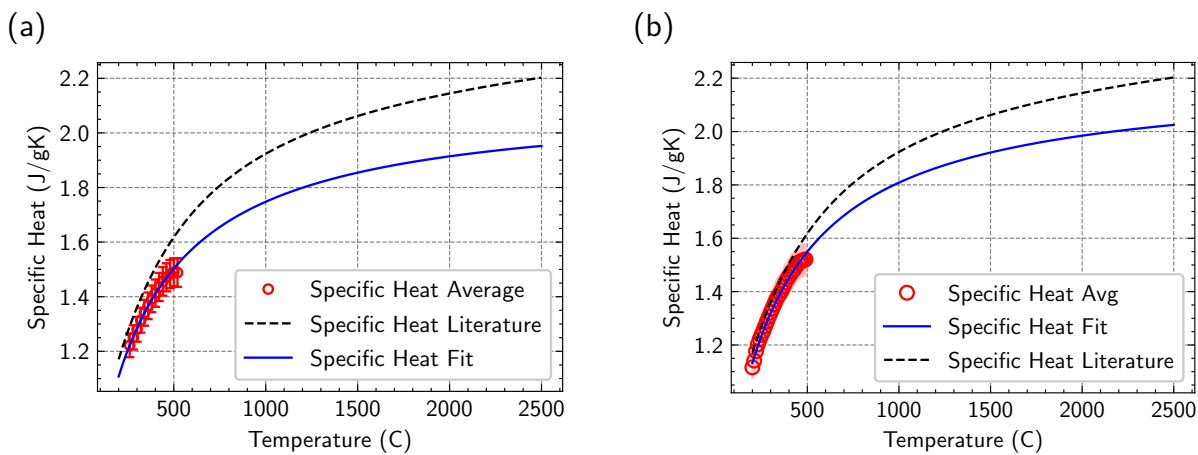

Figure S9: Specific heat measurements for (a) mid-quality graphite and (b) low-quality graphite.

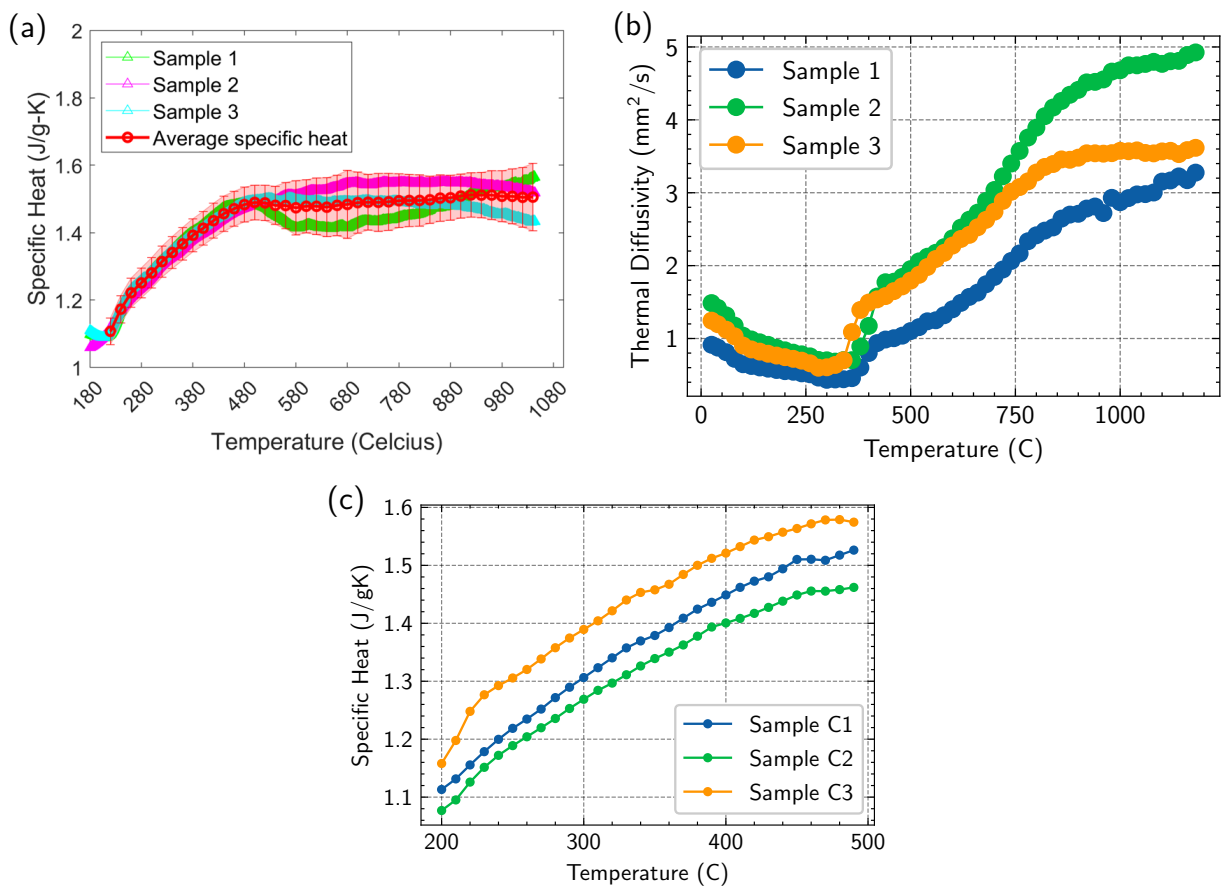

Figure S10: Individual sample measurements for (a) mid-quality graphite specific heat, (b) putty thermal diffusivity, and (c) low-quality graphite specific heat

## 6.2. Steady-state simulations

At the steady-state limit, for techniques such as the guarded hot plate method, we apply a temperature boundary condition on both sides of the sample and measure the heat flux. We then can calculate the effective thermal conductivity using Fourier's law.

In the steady-state case it is relatively straightforward to develop an effective medium theory. We can use the conventional resistance analysis to determine an effective conductivity of the material. In the simplest case, the sample plus coating can be approximated as a series configuration, in a 3-layer model, shown in Figure S11(a). The thickness of putty layer can be determined from the measured added volume of the putty. The effective conductivity can then be calculated as:

$$k_{eff} = \frac{1}{\frac{v_{putty}}{k_{putty}} + \frac{v_{graphite}}{k_{graphite}}}$$

where $v$ is the volume fraction. However, due to the complex interface between the graphite and putty, the 3-layer model may be insufficient. Instead, we introduce a 5-layer model including an effective medium of a mixture of graphite and putty between the sample and coating regions, as shown in Figure S11(b). This allows a better representation of the complex interface sample-coating profile.

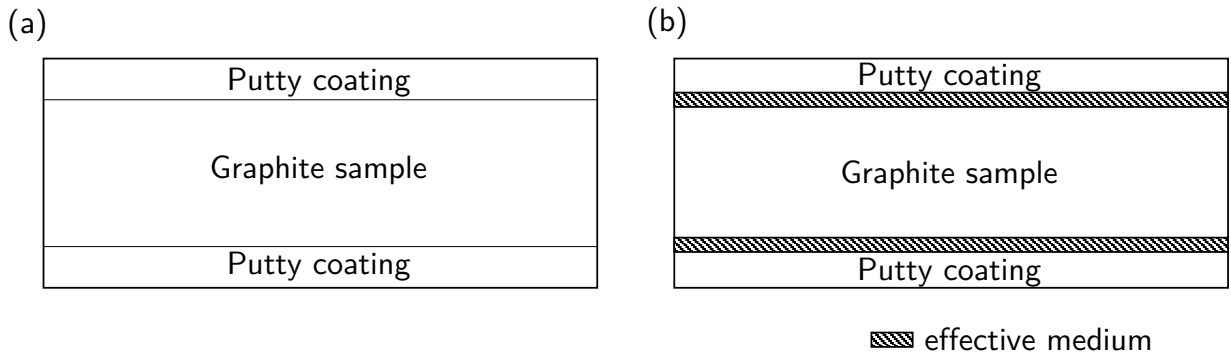

Figure S11: Schematics of 3-layer and 5-layer effective medium models for coated samples. The 3-layer model consists of a simple series approximation considering the volume of putty added. The 5-layer model adds an effective medium layer between the sample and coating to describe the complex interface profile more accurately.

After calculating the effective conductivities of the 100 interface profiles generated, we can determine the form of the interface effective medium that matches results. In this case, the parallel effective medium theory performed the best. Namely, the form of the effective conductivity within the effective medium region was

$$k_{EMT} = v_{putty_{EMT}} k_{putty} + (1 - v_{putty_{EMT}}) k_{graphite}$$

where $v_{putty_{EMT}}$ is the volume fraction of putty in the effective medium region. This depends on the size of the effective medium region, which is defined as

$$V_{EMT} = V_{inside} c_1$$

where $V_{inside}$ is the "missing" area in the graphite volume due to surface porosity and is a measured quantity. $c_1$ is a scaling factor, indicating how large of an area we want the effective medium to encompass. Then, we see that $v_{putty_{EMT}} = 1/c_1$ i.e. if $c_1 = 1$, $V_{EMT} = V_{inside}$ and $v_{putty_{EMT}} = 1$ and we return to the 3-layer approximation. By considering a larger effective medium region we may be able to obtain results more consistent with FEM-simulated values. The final effective medium theory is thus:

$$\frac{L_{total}}{k_{eff}} = \frac{L_{putty}}{k_{putty}} + \frac{L_{graphite}}{k_{graphite}} + \frac{L_{EMT}}{k_{EMT}} \tag{S1}$$

We can tune the value of $c_1$ to minimize error across a wide variety of part geometries and see that $c_1 = 1.05$ accomplishes this, as shown in Figure S12.

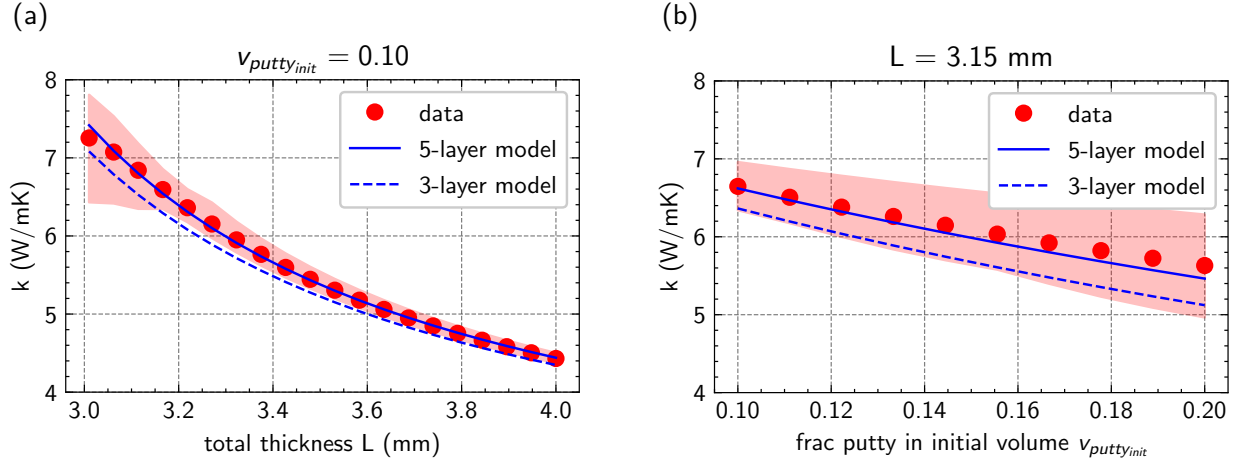

Figure S12: Effective thermal conductivities for (a) varying total part thicknesses and (b) varying fractions of putty in the initial volume. Comparing the 5-layer with the 3-layer model shows the improved accuracy of adding an effective medium layer between sample and coating.

We have thus developed a closed-form effective medium theory to accurately predict effective conductivity from input conductivities (and vice versa) for coated sample materials. While this is useful for steady-state measurement techniques, it is unfortunately not applicable for LFA.

The measured diffusivity is different from the effective or mean thermal diffusivity. The apparent diffusivity is calculated from Equation 1, using the transient temperature response. In contrast, the effective thermal diffusivity can be calculated by using the effective thermal conductivity calculated in Equation S1 with the definition of thermal diffusivity in Equation 2. However, these two values will be different, because the effective thermal conductivity was calculated using the steady-state approximation. Figure 9 further demonstrates this point.

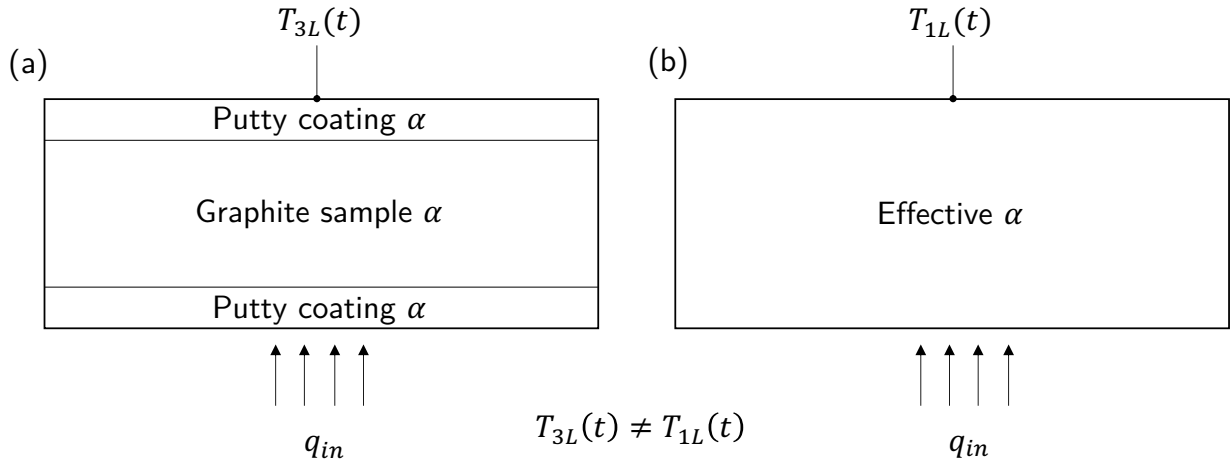

Figure S13: Demonstrating the differences between apparent and effective thermal diffusivity. (a) Shows the real 3-layer coated sample with distinct diffusivities for each layer. (b) Shows the effective diffusivity approximation. Solving the heat equation for (a) and (b) will yield different results, indicating the transient response will be different.